\documentclass[a4paper,10pt]{article}
\usepackage[T1]{fontenc}
\usepackage{graphicx}
\usepackage{booktabs}
\usepackage[misc]{ifsym}

\usepackage[svgnames]{xcolor}
\usepackage{xspace}
\usepackage{siunitx}
\usepackage{nicematrix}
\usepackage{tabularx}
\usepackage{algpseudocode}
\usepackage[algpseudocodecmds,nocitepackage]{pauli_new}
\usepackage{url}

\newcommand{\ownalgo}{\algname{HaPSi}\xspace}
\newcommand{\greedy}{\algname{Greedy}\xspace}
\newcommand{\naive}{\algname{Na\"ive}\xspace}

\newcommand{\wcmam}{\datasetname{Mammals}\xspace}

\newcommand{\dialect}{\datasetname{Dialect}\xspace}
\newcommand{\newsgroups}{\datasetname{20Newsgroups}\xspace}
\newcommand{\abstracts}{\datasetname{Abstracts}\xspace}

\newcommand{\algourl}{https://doi.org/10.5281/zenodo.15629578}

\definecolor{TolHcYellow}{HTML}{DDAA33}
\definecolor{TolHcRed}{HTML}{BB5566}
\definecolor{TolHcBlue}{HTML}{004488}

\begin{document}

\title{Hashing for Fast Pattern Set Selection}

\author{Maiju Karjalainen \and
Pauli Miettinen}

\date{University of Eastern Finland \\ \texttt{firstname.lastname@uef.fi}}

\maketitle              %

\begin{abstract}
  Pattern set mining, which is the task of finding a good set of
  patterns instead of all patterns, is a fundamental problem in data
  mining. Many different definitions of what constitutes a good set
  have been proposed in recent years. In this paper, we consider the
  reconstruction error as a proxy measure for the goodness of the set,
  and concentrate on the adjacent problem of how to find a good set
  efficiently. We propose a method based on bottom-k hashing for
  efficiently selecting the set and extend the method for the common
  case where the patterns might only appear in approximate form in the
  data. Our approach has applications in tiling databases, Boolean matrix factorization, and redescription mining, among others. We show that our hashing-based approach is significantly faster than the standard greedy algorithm while obtaining almost equally good results in both synthetic and real-world data sets.

\end{abstract}

\section{Introduction}
\label{sec:introduction}

The goal of data mining is to find surprising, new information about the data. A common problem with many methods that find local patterns from the data is that they find \emph{too many} of those patterns---so-called pattern explosion problem. The popular solution to the pattern explosion problem is to switch to \emph{pattern set mining}; that is, to the task of finding a good set of patterns instead of all patterns. 

The definition of what is a good set is a topic of active study, and approaches such as MDL~\cite{vreeken2011} or subjective interestingness~\cite{debie2011} have been proposed for that. In this paper we approach the problem from a different perspective. We take the reconstruction error as the means of defining a good set, and ask the question: how fast can we find the set of patterns that minimizes the reconstruction error?

The motivation for our approach is twofold: For one, we argue that the reconstruction error is often a good proxy for the more sophisticated pattern set selection methods. Indeed, they all ask the question: how much yet unexplained data this new pattern explains, the main difference being how the ``how much'' part is defined. Of these, the ``plain'' reconstruction error is also the fastest to compute, which is why we prefer it. 

For two, we argue that selecting the set of patterns is typically slow. This can be shadowed by the fact that \emph{generating} the patterns can be slower and hence speeding up the task of selecting the patterns does not feel important. But this is often not so straightforward. In many cases, each pattern is independent of the others, and they can be generated efficiently in parallel. With modern multi-core CPUs and GPUs the pattern generation, even for a huge number of patterns, can be very effective. \emph{Selecting} the patterns in the set, however, is inherently sequential process: in a good pattern set every selected pattern depends on the others. The selection of the patterns can easily become the slower part of the process.

To obtain our fast pattern set selection algorithm, we utilize the bottom-$k$ hashing idea~\cite{cohen2007summarizing,cohen1997size,thorup2013bottom} using the efficient computation proposed in~\cite{amossen14better}. While~\cite{amossen14better} studies, in some sense, the quality of a pattern set, our approach will build the set, and furthermore, we propose a way to handle inexact patterns, that is, patterns that do not appear as such in the data.

In the next section we will formally define our problems and analyze their computational complexity. We will also discuss how our problems model different pattern set mining problems, although we postpone the other related work to Sect.~\ref{sec:related-work}. We will end Sect.~\ref{sec:notation-problems} with a brief analysis of the standard greedy algorithm before introducing our algorithm in Sect.~\ref{sec:algorithms}. The experiments are presented in Sect.~\ref{sec:experiments}.

\section{Notation, Problems, and Theoretical Analysis}
\label{sec:notation-problems}

In this paper, we study the problem of selecting a good subset of patterns to explain the data. There are two variants: in \emph{exact patterns} the patterns are exact subsets of the data; in \emph{inexact patterns} the patterns can cover things that are not in the data. We formalise the problems using binary matrices. For that, we need the following concepts. We denote a matrix (Boolean or otherwise) with a bold-face upper-case letter, such as $\mM$. Vectors are denoted with bold-face lower-case letters, such as $\vv$. All vectors are considered to be column vectors. The $(i,j)$ element of matrix $\mM$ is $m_{ij}$ and the $i$th element of vector $\vv$ is $v_i$.  We say that a binary matrix $\mS\in\B^{m\times n}$ is \emph{dominated by} a binary matrix $\mD\in\B^{m\times n}$ if $s_{ij} \leq d_{ij}$ for all $i$ and $j$. We denote that by $\mS\leq\mD$. Given a collection of binary matrices $\col{C} = \{\mS_i : i=1,\ldots, \ell, \mS_i\in\B^{m\times n}\}$, we define $\bigvee_{\mS\in\col{C}}\mS = \mC$ as the element-wise logical \emph{or} over matrices $\mS$, that is, $c_{ij} = \bigvee_{\mS\in\col{C}}s_{ij}$. Matrix $\mM$, of size \by{m}{n}, is \emph{rank-1} if it can be expressed as an outer product of two vectors, $\mM = \vu\vv^T$. Matrix $\mM\in\B^{m\times n}$ is \emph{Boolean rank-1} if $\mM=\vu\vv^T$ where $\vu\in\B^m$ and $\vv\in\B^n$. Finally, we use $\abs{\cdot}$ to denote the number of non-zero entries in a matrix or a vector, that is, $\abs{\mM} = \abs{\{(i,j) : m_{ij} \neq 0\}}$, where the latter $\abs{\cdot}$ is the usual set cardinality operator. 

\subsection{Problem Statement}
\label{sec:problem-statement}

We give the exact definitions of our problems below. For an explanation of how these definitions relate to pattern set mining tasks, see Section~\ref{sec:conn-other-forms}.

\begin{problem}[Exact Patterns]
  \label{prob:exact}
  Given a binary data matrix $\mD\in\B^{m\times n}$, a collection of (rank-1) binary matrices of the same size, $\col{S}=\{\mS_1, \mS_2, \ldots, \mS_{\ell} : \mS_i\in\B^{m\times n}, \mS_i\leq \mD\}$, and an integer $k$, find a subcollection $\col{C}\subseteq\col{S}$ of size $k$ that minimizes
\begin{equation}
  \label{eq:max_cover}
  \abs*[Big]{\mD - \bigvee_{\mS\in\col{C}}\mS}\; .
\end{equation}
\end{problem}

Alternatively, the size of collection $\col{C}$ can be unlimited and \eqref{eq:max_cover} can be replaced with a requirement that
\begin{equation}
  \label{eq:exact_cover}
  \bigvee_{\mS\in\col{C}}\mS = \mD\; ,
\end{equation}
where we now assume that $\cup\col{S} = \bigcup_{\mS\in\col{S}}\mS = \mD$. Notice that as $\mS$ are dominated by $\mD$, the union can never have 1s in locations where $\mD$ does not have them. Consequently, the subtraction in \eqref{eq:max_cover} cannot produce negative entries. We say that $\cup\col{S}$ does not \emph{cover} any $0$s of $\mD$.

When the pattern matrices are not dominated by the data, the problem definition is as follows.

\begin{problem}[Inexact Patterns]
  \label{prob:inexact}
  Given a binary data matrix $\mD\in\B^{m\times n}$ and a collection of (rank-1) binary matrices of the same size, $\col{S}=\{\mS_1, \mS_2, \ldots, \mS_{\ell} : \mS_i\in\B^{m\times n}\}$, find a subcollection $\col{C}\subseteq\col{S}$ that minimizes
  \begin{equation}
  \label{eq:inexact_cover}
  \abs*[Big]{\mD - \bigvee_{\mS\in\col{C}}\mS}\; .
\end{equation}
\end{problem}

Contrary to \eqref{eq:max_cover}, in \eqref{eq:inexact_cover} the matrices $\mS$ are not necessarily dominated by $\mD$ and hence the subtraction can yield negative values. Similarly to Problem~\ref{prob:exact}, it is also possible to limit the size of $\col{C}$ in Problem~\ref{prob:inexact} by some user-defined constant $k$.

In the above definitions, we have limited the patterns to be rank-1 binary matrices. As we shall see below, this still allows us to represent many different kinds of patterns commonly encountered in data mining. Our algorithm will utilize the rank-1 structure for efficiency; however, we can still handle arbitrary binary matrices as patterns, with a cost in the running time of the algorithm (see Sect.~\ref{sec:estimatingtilesize}). We will use terms \emph{pattern}, \emph{tile}, and \emph{rank-1 matrix} interchangeably in this paper.

\subsection{Computational Complexity}
\label{sec:comp-compl}

Both problems (and their variants) are $\NP$-hard. The Max $k$-Cover problem can be reduced to Problem~\ref{prob:exact}, and the Set Cover problem to the variant of Problem~\ref{prob:exact} with unlimeted cover size (see, e.g.~\cite{geerts2004a}). The simplest form of this reduction is to consider a case where $\mD$ has only one $m$-dimensional column full of $1$s. Each $\mS$ is also an $m$-dimensional column vector and they correspond to incidence vectors of the sets in Set Cover or Max $k$-Cover. It is trivial to see that solutions that maximize~\eqref{eq:max_cover} or \eqref{eq:exact_cover} maximize Max $k$-Cover or Set Cover, respectively.

In the Positive--Negative Partial Set Cover~\cite{miettinen2008}, the input consists of positive and negative elements and a collection of sets of those elements and the goal is to select a subcollection of the sets that minimizes the sum of covered negative elements plus the sum of uncovered positive elements. Consider again a case where $\mD$ is $m$-dimensional vector. Each entry $d_i$ of the input vector corresponds to an element in the input; $d_i=1$ if the element is positive and $d_i=0$ otherwise. The matrices $\mS$ are again incidence vectors of the sets. It is straightforward to see that minimizing~\eqref{eq:inexact_cover} is the same as minimizing the sum of uncovered $1$s plus the sum of covered $0$s.

Furthermore, these reductions are approximation-preserving as the values of cost functions (number of used matrices/sets or number of uncovered $1$s plus covered $0$s/uncovered positive and covered negative elements) are exactly the same after the reductions. Hence, Problem~\ref{prob:exact} with optimization goal~\eqref{eq:max_cover} cannot be approximated better than $\Omega(1 - 1/e)$. With optimization goal~\eqref{eq:exact_cover} the approximation lower bound is $\Omega(\ln(m))$~\cite{feige1998}. For Problem~\ref{prob:inexact}, the approximation lower bound is much higher $\Omega(2^{\log^{1-\varepsilon}\ell^4})$, where $\ell$ is the number of sets in $\col{S}$~\cite{miettinen2008}. The intuition behind the vast difference in inapproximability lower bounds is that in the case of exact patterns, we should aim to select maximally disjoint patterns, whereas in the case of inexact patterns, we should aim to select patterns that cover maximally disjoint set of $1$s but maximally similar set of $0$s, as we only ``pay'' the cost of covering a $0$ once. This latter problem is clearly much harder.

\subsection{Connections to Other Forms of Pattern Set Mining}
\label{sec:conn-other-forms}

While we described our problems in the terms of binary matrices, it is relatively easy to see that the description applies to many different pattern set mining tasks. Here we describe three such tasks: tiling databases, Boolean matrix factorization, and redescription mining.

\emph{Tiling databases}~\cite{geerts2004a} is the task of selecting a subset of (frequent or closed) itemsets to cover all the items in a transaction database. In our setting, the transaction database is the data matrix $\mD$ with items as columns and transactions as rows, and each rank-1 matrix is one itemset: the columns with $1$s correspond to the items in the itemset and the rows with $1$s correspond to the transactions that contain the itemset. As every item of the itemset must appear in a transaction for the transaction to contain the itemset, the patterns are dominated and we have the exact patterns of Problem~\ref{prob:exact}.

Equation~\eqref{eq:max_cover} is a natural optimization goal for tiling. Equation~\eqref{eq:exact_cover} requires that every item in every transaction is included in some pattern; this is achieved most easily by including the singleton itemsets for each item in the set $\col{S}$. Such exact cover is also required when one uses alternative means of selecting the pattern set, such as MDL~\cite{vreeken2011} (see also Sect.~\ref{sec:related-work}).

\emph{Boolean matrix factorization}~\cite{miettinen2020} is the task of (approximately) expressing a given binary matrix as a Boolean product of two binary factor matrices. Equivalently, it is the task of expressing the input as a element-wise Boolean \emph{or} of rank-1 binary matrices. With this definition, the connection to our problem is clear. As Boolean factorization can be approximated both by not covering some $1$s and by covering some $0$s, the appropriate formulation is typically that of Problem~\ref{prob:inexact}. For computing the smallest exact Boolean factorization (so-called \emph{Boolean rank}~\cite{monson1995}), however, Problem~\ref{prob:exact} with optimization goal~\eqref{eq:exact_cover} is the appropriate formulation.

\emph{Redescription mining}~\cite{ramakrishnan2004,galbrun2018a} is a data analysis method that, unlike the aforementioned ones, can, and often does, use numerical data. The goal of redescription mining is to mine \emph{redescriptions}, that is, pairs of queries $(q_L, q_R)$ over two sets of data, $\mD_L$ and $\mD_R$, that provide two different views to the same entities. A classical example of redescription mining is species presence--absence data and climate data. These provide different ``views'' to the same geographical locations and a good redescription on such data would describe a query (or a rule) over the species (i.e.\ the presence or absence of certain species) and a query over the climate variables (e.g.\ temperature or precipitation) such that the two queries hold in approximately same geographic areas. The quality of a redescription is often measured using the Jaccard coefficient between the sets of entities where the individual queries hold. 

\begin{figure}[tbp]
  \includegraphics{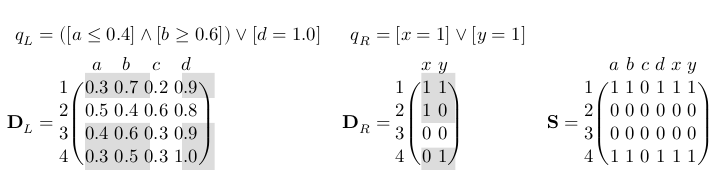}
  \caption{Example of how redescriptions can be expressed in the framework of Problem~\ref{prob:exact}. \textbf{Top:}  A pair of queries, forming a redescripition. \textbf{Bottom left and middle:} Original data matrices $\mD_L$ and $\mD_R$ for redescription mining. Cells shaded in gray will be $1$ in $\vu_L\vv_L^T$ and $\vu_R\vv_R^T$. \textbf{Bottom right:} The final matrix $\mS = (\vu_L\land\vu_R)[\vv_L^T, \vv_R^T]$.}
  \label{fig:example}
\end{figure}

While the data in redescription mining can be numerical, every individual query can be seen as a rank-1 binary matrix. The $1$s in the columns correspond to the attributes that appear in the query; the $1$s in the rows correspond to the entities where the query holds. For a pair of queries, we can concatenate these two matrices next to each other (note that they must have the same number of rows) and consider only those rows where both matrices have a $1$. That is, if $\vu_L\vv_L^T$ is the rank-1 matrix corresponding to $q_L$ and $\vu_R\vv_R^T$ is that for $q_R$, the rank-1 matrix corresponding to pair $(q_L, q_R)$ is $(\vu_L\land\vu_R)[\vv_L^T, \vv_R^T]$, where $\land$ is the element-wise \emph{and} and the right-hand vectors are concatenated. Figure~\ref{fig:example} shows an example of one pair of queries over numerical and Boolean data and the resulting pattern matrix $\mS$.

\subsection{The Greedy Algorithm}
\label{sec:greedy-algorithm}

By far the most common algorithm to do the pattern set selection for minimum error case is the greedy algorithm. This algorithm takes first the pattern that minimizes the error the most, then the pattern from the remaining ones that minimizes the error the most given the already selected pattern, and so on and so forth until it has either selected a predefined number of patterns or no further pattern reduces the error. This approach is used in tiling~\cite{geerts2004a}, in BMF~\cite{miettinen2020}, and in redescription mining\footnote{In redescription mining the approaches in literature have some technical differences to the approach outlined above.}~\cite{galbrun2018a} for selecting a set of patterns. For the exact case, the greedy algorithm has approximation ratio of $O(1 - 1/e)$ (for~\eqref{eq:max_cover}) or $O(\ln(m))$ (for~\eqref{eq:exact_cover}), matching the lower bounds. For inexact patterns, however, the greedy algorithm can be arbitrarily bad, although it seems to work well in practice~\cite{miettinen2020}.

The time complexity of the greedy algorithm is $\tilde{O}(mn\ell^2)$ for data matrix of size \by{m}{n} and $\abs{\col{S}} = \ell$: after selecting every new pattern, the algorithm must re-compute how good the remaining patterns are. We use $\tilde{O}(\cdot)$ to denote the time complexity that ignores the logarithmic factors (that are needed to find the best pattern after the updates). For sparse patterns, tighter analysis can be obtained, as updating each pattern takes at most as much time as the number of $1$s in the pattern.

\section{The Algorithm}
\label{sec:algorithms}

Our algorithm, called \ownalgo (Hashing for Pattern Selection), is based on using hash values to estimate the sizes of rank-1 matrices and their combinations. The main idea, presented in \cite{bar2002counting} and \cite{amossen14better}, is to calculate hash values uniformly distributed in $[0,1]$ for all $1$s in a rank-1 matrix, and to use the $k$th smallest hash value $v$ to estimate the number of $1$s in the rank-1 matrix as $k/v$. This can be used as an estimate, because, if there are $z$ values uniformly distributed in $[0,1]$, the expected value of the $k$th smallest value is $k/(z+1)$. This method is used by \cite{amossen14better} to estimate the product of two Boolean matrices, by iterating efficiently over the rank-1 matrices and keeping track of the $k$ smallest hash values seen at each point.\!\footnote{Notice that this $k$ is different to the number of selected tiles in problem definitions.} Next we will explain a technique for estimating sizes of overlapping tiles before explaining the algorithm in Sect.~\ref{sec:exactalgorithm}. We analyse the time complexity in Sect.~\ref{sec:timecomplexity}.

\subsection{Estimating Overlapping Tile Size}
\label{sec:estimatingtilesize}

This section describes a hashing-based technique of estimating the number of $1$s in a single tile and in a collection of overlapping tiles. It is based on \cite{amossen14better}, although with some changes to make the approach suitable for our application.

The hash value $h(x,y)$ for a point $(x,y)$ in a rank-1 matrix is computed based on two pairwise independent hash functions $h_1, h_2\colon U \rightarrow [0,1]$ which map the row and column indices to $[0,1]$. The hash value for the point is then defined as $h(x,y) = (h_1(x) - h_2(y))\mod 1$.

To find the $k$ smallest hash values of a rank-1 matrix $\vu\vv^T$, $h(x,y)$ are arranged into a matrix of size $\by{\abs{\vu}}{\abs{\vv}}$. The rows are ordered by increasing values of $h_1$ and the columns by increasing values of $h_2$. Going down a column the values of $h(x,y)$ will increase, except at one point where the value drops to the minimum value of that column. Similarly on the rows, the values decrease going to the right, except for one increase to the maximum value of that row. 

The algorithm is given a value $p$ which serves as an initial maximum hash value to be stored. The values of the hash matrix are visited starting from the top of the left-most column, first finding the smallest value of the column. After that, going through the column, all values that are smaller than $p$ are stored. After all such values have been found, the next column is searched, this time starting one row below the row where the previous column had its smallest value.
When $k$ values have been found, if the $k$th value is smaller than $p$, $p$ is updated to this new smallest value.

The idea of using the $k$th smallest hash value as a size estimate can be applied to a matrix of any rank, but the efficiency of this algorithm comes from iterating over the row and column indices which only works on rank-1 matrices. Hence, if we want to use non-rank-1 patterns, we have to give up on this optimization.

\subsection{The Algorithm for the Exact Patterns}
\label{sec:exactalgorithm}

The first step of our algorithm is to compute the hash values for each tile. This is done as explained above. For pairwise independent hash functions, we use random affine transformations in $\Z/p$, where $p\in\N$ is a prime number larger than largest dimension of the matrix. That is, each hash function is of form $h(x) = (ax + b\mod p) / p$ for some random $a$ and $b$. Unlike \cite{amossen14better}, we store the $k$ smallest hash values separately for each tile instead of maintaining a single collection of size $k$ over all tiles.

The pseudocode for \ownalgo is given in Algorithm~\ref{alg:ownalgo}. The algorithm starts by choosing the largest tile. Then, it goes through rest of the tiles, trying to find the one that increases the number of ones covered the most. For each tile that has not been chosen yet, we concatenate its hash values and those of the already-chosen tiles' in line~\ref{alg:combinedhashes}. The combined hash values are sorted and we use the $k$th smallest value to get the estimated size of these tiles together. In our preliminary experiments we noticed that this method gives generally very accurate estimates, but sometimes gives an exceptionally large or small result. This is why we get the final size estimate as the median over several ($\abs{\col{H}}$, typically $10$ in our experiments) estimates (line~\ref{alg:tilecontribution}).

\begin{algorithm}[tb]
    \caption{Exact Patterns (Problem 1)}\label{alg:ownalgo}
    \begin{algorithmic}[1]
      \Input Data matrix $\mD$, tiles $\col{T}$, collection of vectors of hashvalues for each tile $\col{H}$, maximum number of tiles returned $t_{\max}$, $k$ for bottom-$k$ hashing, and maximum number of tiles searched in each iteration $m$
      \Output A set of tiles $\col{Q}$
      \Function{\ownalgo}{T,H,t} 
      \State $\col{Q}\gets$ tile $t_0$ with the smallest reconstruction error
      \State $\col{T}\gets \col{T}\setminus t_0$
      \While{$\operatorname{len}(\col{Q}) < t_{\max}$}
        \State $\col{C}\gets\emptyset$
          \For{ $t\in \col{T}$}
            \State $\col{E}\gets \emptyset$ 
            \For{$\vh \in \col{H}[t]$}
                \State $\vh \gets \Call{Sort}{\col{H}[\col{Q}] \cup \vh}[:k]$\label{alg:combinedhashes}
                \State $\col{E}\gets\col{E}\cup \{k/\vh[k]\}$
            \EndFor
            \State  $v \gets \operatorname{median}(\col{E})$ \label{alg:tilecontribution}
            \State $\col{C}\gets\col{C}\cup \{(v,t)\}$ 
          \EndFor
          \State  $\col{C}\gets \Call{Sort}{\col{C}}$ in descending order by $v$ 
          \For{$(v,c) \in \col{C}[:m]$} \label{alg:lookform}
            \If {$\operatorname{error}(\col{Q}\cup \{c\})< \operatorname{error}(\col{Q})$}
                \State $\col{Q}\gets\col{Q} \cup \{c\}$ 
                \State \textbf{break}
            \Else
                \State \textbf{return} $\col{Q}$
            \EndIf
          \EndFor
      \EndWhile
    \State \textbf{return} $\col{Q}$
    \EndFunction
    \end{algorithmic}
  \end{algorithm}

When we have the estimated sizes for all tiles, we sort the tiles in a descending order according to the estimate. Then, in line~\ref{alg:lookform}, we go through a limited number $m$ (typically $30$) of top candidates, and calculate the actual reconstruction error (or the number of $1$s in the resulting matrix in the exact case) we would get when adding that tile to the already chosen ones. Once we find a tile that lowers the reconstruction error, we stop the search and add that tile to the collection. The process is repeated until the error no longer improves, or the maximum number of tiles is met (constraint set by the user).

\paragraph{Handling inexact tiles.}
In the case where the tiles can cover zeros, the first tile chosen is the one with the smallest reconstruction error, meaning it has the least amount of uncovered ones and the least amount of covered zeros. In addition to estimating the combined size with the already chosen tiles, we subtract the number of zeros covered by the tile from the joined size estimate. This means that the line \ref{alg:tilecontribution} changes to $\operatorname{median}(\col{E}) - \col{Z}_t$, where $\col{Z}_t$ is the number of zeros the tile $t$ covers. %

We note that by subtracting every $0$ from every tile that covers it, we are probably over-penalizing the tiles; if an already-chosen tile covers some of the same $0$s as this tile, using this tile does not increase the cost for those $0$s. While this approach might seem na\"ive, the inapproximability of the inexact case indicates that much better algorithms are unlikely to exist. 

\subsection{Time Complexity}
\label{sec:timecomplexity}

For an input matrix of size \by{m}{n} and number of tiles $\ell$, the time complexity of \ownalgo is $\tilde{O}((n+m)\ell+\ell^2 + mn\ell)$. Computing the hash values for one tile $\vu\vv^T$ takes amortized time $O(\abs{\vu}+\abs{\vv})$~\cite{amossen14better}, where $\abs{\vu}$ is the number of non-zeros in $\vu$. This explains the first term. After each tile has been chosen, we need update how well the others do given the chose tile (the second term), and to choose the tile, we need to compute its exact error (the final term). We could omit calculating the actual error, but we would then be susceptible to cascading errors in size estimates.

\section{Experiments}
\label{sec:experiments}

We evaluate the performance of our algorithm on both synthetic and real-world data. The aim of the synthetic data experiments is to assess how different data features affect the algorithm; experiments with real-word data then validate these results. 

\subsection{Experimental Setup}
\label{sec:experimental-setup}

We generate the synthetic data as follows. First, we make \num{100} random rank-1 matrices (aka tiles or patterns) of size \by{m}{n} by sampling vectors $\vu_i\in\B^m$ and $\vv_i\in \B^n$ from $m$- and $n$-dimensional Bernoulli distributions with parameter $p\in (0,1)$. Each ``original'' tile is $\vu_i\vv_i^T$. We make 5 copies of each of these tiles, and for each copy $\vu_i'\vv_i'^T$, we move a fraction $d$ of the $1$s from $\vu_i'$ and $\vv_i'$ to other locations. This gives us \num{100} mostly non-overlapping rank-1 matrices, each of which have \num{5} copies that have a given amount overlap with the original and each other.

Then, we have two ways of creating a data matrix out of these \num{600} patterns, either by multiplying the original \num{100} patterns, or by multiplying all \num{600}. After this we can add noise to the data matrix by randomly flipping a portion of the elements.

For the exact pattern experiments we only created the data matrix by multiplying all \num{600} tiles and did not add any noise, to ensure that the tiles are dominated by the data. For the inexact pattern experiments, we used both methods of creating the data matrix and also added noise.

There are four features we test with the synthetic matrices: overlap between the tiles, density of the data matrix and the tiles, size of the matrix, and the added noise. We tested all of these with the inexact patterns, and the first two (overlap and density) also with the exact patterns.

We created the data \num{5} times for each set of parameters tested, and the results are the average over these five runs. As our own algorithm has some randomness in creating the hash values, we initially tested it with three restarts on the same data. However, since the algorithm uses the median estimate over multiple hash values on a single run, running it more times did not have much variance in the results.

We compared \ownalgo to two baselines. \greedy is the standard greedy algorithm for Set Cover (see Sect.~\ref{sec:greedy-algorithm}). \naive is a simpler and faster algorithm that always selects the tile from the unselected tiles that alone covers the most $1$s (or reduces the error most, in inexact case), irrespective of what has been selected before. These two algorithms represent two extremes. \greedy, while not being optimal, is the state of the art solution for minimizing the error, while \naive is about as fast as it can get: it's time complexity is $\tilde{O}(nm\ell)$, that is, it only requires one linear scan over the input data.

For all experiments, we report \emph{relative reconstruction error}, that is, reconstruction error divided by the matrix size. 
The algorithms were implemented on Python and are freely available.\footnote{\algourl} The experiments were conducted on a machine with 2 AMD EPYC 7702 processors with 64 cores each. All experiments were run single-threaded.

\subsection{Setting the Hyperparameters for \ownalgo}
\label{sec:hyperparameters}

We evaluated the effect of the hyperparameters of \ownalgo on the reconstruction error and the running time. The following hyperparameters were tested: number of hash values used in bottom-$k$ hashing, $k$; number of times the hashing is repeated for the data, $\abs{\col{H}}$; and the maximum number of tiles for which the reconstruction error is computed before stopping, $m$. The other parameter for Algorithm \ref{alg:ownalgo} is the maximum number of tiles returned, $t_{\max}$, but it was not tested here since it only affects the final error and running time.

The default values used for the experiments were $m=30$, $k=30$, $\abs{\col{H}}=10$. The tests were ran on a synthetic data of size \by{1000}{1200} with data density \num{0.06}; fraction of rows and columns changed in copied tiles by \num{0.1} and noise added to the data matrix was \qty{10}{\percent}. Each hyperparameter was varied separately, while keeping others constant. 

\begin{figure}[tbp]
  \centering
  \includegraphics{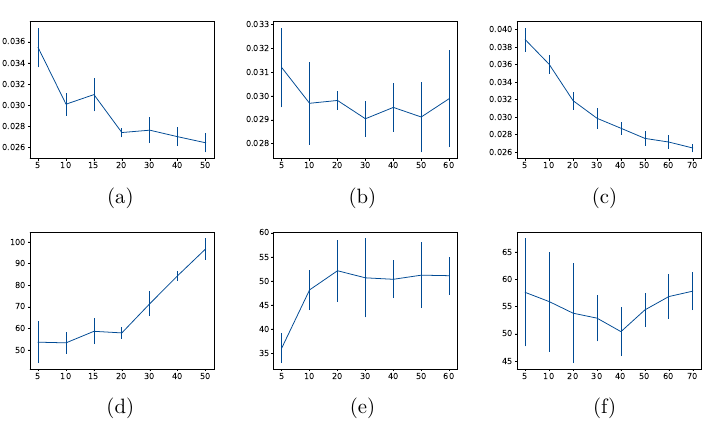}
    \caption{Reconstruction errors and running times from the hyperparameter tests. \textbf{Top row:} $x$-axis: (a) $\lvert\col{H}\rvert$,
      number of times the hashing is repeated; (b) $m$, maximum number of tiles for which the reconstruction error is computed; (c) $k$, number of hash values. Relative reconstruction error on $y$-axis. \textbf{Bottom row:} Running time in seconds on $y$-axis. (d)--(f) as (a)--(c).}
    \label{fig:hyperparameters}
  \end{figure}

The results of the tests are shown in Fig.~\ref{fig:hyperparameters}. The results show that two of the parameters had a noticeable effect on the reconstruction error: $\abs{\col{H}}$, the number of times the hashing was repeated, and the $k$ for the bottom-$k$ hashing method (Fig.~\ref{fig:hyperparameters}(a) and (c)). Naturally as the hash values are computed several times, the final estimate (calculated as the median over all of the estimates) will be more accurate. A larger value here also significantly affects the running time. The choice of optimal value of $k$ is explained in detail in \cite{amossen14better}, and our results show that a larger value improves the error at least to some point. However, we chose not to use a large value of $k$ in our experiments, as in real-world applications the rank-1 matrices can have varying densities; in particular, rank-$1$ matrices with less than $k$ $1$s would be automatically discarded. Another approach for choosing the value of $k$ would be to choose it after all of the rank-$1$ matrices have had their hash values calculated, and their sizes are known, but testing this is left for future work.

\subsection{Results on Synthetic Data}
\label{sec:results-synth-data}

\begin{figure}[tbp]
  \centering
  \includegraphics{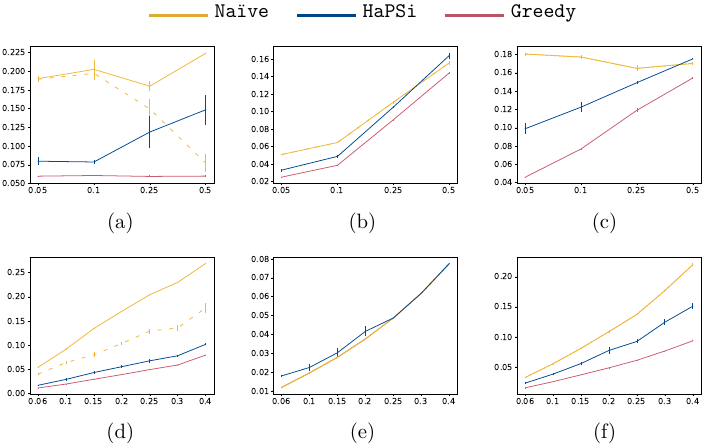}
    \caption{Relative reconstruction error on $y$-axis. \textbf{Top row:} The amount $d$ of changed $1$s on $x$-axis. (a) Inexact case with \num{100} tiles. (b) Inexact case with \num{600} tiles. (c) Exact case with \num{600} tiles. In (a), the dashed line indicates the best error \naive obtained. \textbf{Bottom row:} Data density on $x$-axis. (d)--(f) as (a)--(c).}
    \label{fig:overlap-density}
  \end{figure}

In each test, all other data generation parameters except the one being tested for were kept constant. The default values used in the data generation were as follows. Data size: \by{1000}{1200}; data density: \num{0.3}; fraction of rows and columns changed in copied tiles: \num{0.1}; noise added to the data matrix: \qty{10}{\percent}. Each algorithm searched for a maximum of \num{200} tiles unless otherwise specified.

\paragraph{Overlap between the tiles.} We created the data as explained above. In the step where the \num{100} tiles are copied, we vary the fraction of rows and columns that are changed. The fractions we used were \numlist{0.05;0.1;0.25;0.5}. 

Reconstruction errors for these experiments are shown in Fig.~\ref{fig:overlap-density}(a)--(c). There the overlap \emph{decreases} as $y$-axis increases. \naive can include tiles that increase the error. With \num{100} tiles, we also show the best result it obtained. This is depicted as a dashed line in Fig.~\ref{fig:overlap-density}(a).

With \num{600} tiles, the results show that the error increases as the overlap reduces. This is most likely due to the fact that there are more $1$s to be covered as each tile is more disjoint. In case of \num{100} tiles, situation is different. Now, there are basically \num{100} ``correct'' tiles and \num{500} ``incorrect'' tiles to select, with increasing non-overlap making the incorrectness worse. Here we see that \greedy always selects the correct tiles. \ownalgo selects good tiles initially, but starts deteriorating as the tiles get more disjoint. \naive is always the worst eventually, but its best result is actually better than \ownalgo with least overlap.

\paragraph{Data density.} We considered data with density \numlist{0.06;0.1;0.15;0.2;0.25;0.3;0.4}. The tile density was adjusted depending whether \num{100} or \num{600} tiles were used, so that the data matrix has the desired density. The results are depicted in Fig.~\ref{fig:overlap-density}(d)--(f). All algorithms perform worse as density increases with \ownalgo being close to \greedy in all cases and \naive being clearly worse in Fig.~\ref{fig:overlap-density}(d) and (f), even when considering its best result in (d). 

\begin{figure}[tb]
  \centering
  \includegraphics{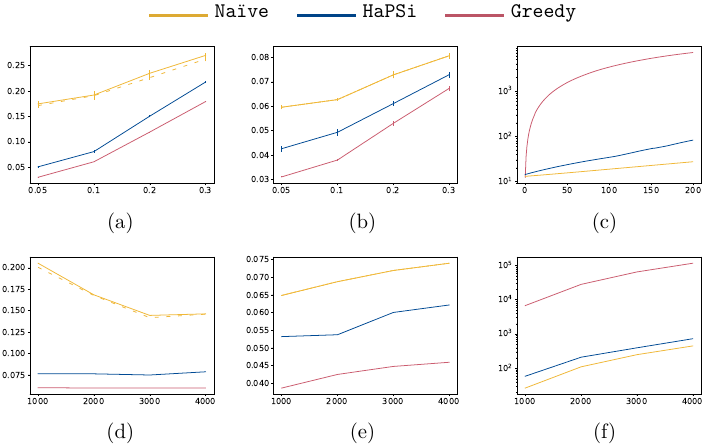}
    \caption{\textbf{Top row, (a)--(b):} The amount of noise on $x$-axis, relative reconstruction error on $y$-axis. (a) Inexact, \num{100} tiles. (b) Inexact, \num{600} tiles. In (a), the dashed line indicates the best error \naive obtained. \textbf{Top row, (c):} Running time (seconds, logarithmic scale) as number of selected patterns increases. \textbf{Bottom row, (d)--(e):} Number of rows in data on $x$-axis, relative reconstruction error on $y$-axis. (d)--(e) as (a)--(b). \textbf{Bottom row, (f):} Running time (seconds, logarithmic scale) as matrix size increases.}
    \label{fig:noise-size}
  \end{figure}

\paragraph{Added noise.} We added \qtylist{5;10;20;30}{\percent} noise to the data matrix. Results are shown in Fig.~\ref{fig:noise-size}(a)--(b). Again, as noise increases all methods get worse, and the order is clear: \naive (worst), \ownalgo, and \greedy, with \ownalgo being relatively close to \greedy.

  \paragraph{Data size.} Finally we tested data size by generating matrices of size \by{x}{(x+200)} for $x$ values of \numlist{1000;2000;3000;4000}. The reconstruction errors are depicted in Fig.~\ref{fig:noise-size}(d)--(e), showing that \ownalgo performs very similarly to \greedy in terms of reconstruction error.

  \paragraph{Running time.} Running times of the algorithms with respect to number of tiles and size of the data are shown in Fig.~\ref{fig:noise-size}(c) and (f), respectively. These plots show clearly that \greedy is significantly slower that \ownalgo, which in turn is very close to \naive, showing that \ownalgo reaches a good balance between result quality and running time.

\subsection{Results on Real-World Data}
\label{sec:results-real-world}

\paragraph{Data sets.}

We used four real-world data sets in our experiments. Some statistics about them are presented in Table~\ref{tab:real-data}. The \wcmam dataset contains information about which mammal species inhabit which areas of the world on one side, and climate information on the other side~\cite{hijmans2005}. The \dialect data contains features of spoken dialects of Finnish over different geographical regions~\cite{embleton1997,embleton2000}. The \newsgroups data\footnote{\url{https://archive.ics.uci.edu/dataset/113/twenty+newsgroups}} is a corpus of \num{1000} posts from \num{20} different newsgroups, stemmed and with rare terms removed. The \abstracts data\footnote{\url{https://archive.ics.uci.edu/dataset/134/nsf+research+award+abstracts+1990+2003}} is another corpus, this time of project abstracts. Terms are stemmed and rare ones are removed. 

\begin{table}[tbp]
  \caption{Properties of real-world data sets}
  \label{tab:real-data}
  \centering
\begin{tabular}[tbp]{@{}lS[table-format=5.0,group-minimum-digits=3]S[table-format=5.0,group-minimum-digits=3]S[table-format=2.2]p{3cm}@{}}
  \toprule
  Data & {Rows} & {Columns} & {Density (\unit{\percent})} & \multicolumn{1}{c}{Note} \\
  \midrule
  \wcmam & 54013& 4802 & {---} & Numerical data \\
  \dialect & 1334 & 506 & 16.14 & \\
  \newsgroups & 5163 & 19997 & 0.89 & \\
  \abstracts & 4894 & 12841 & 0.90 & \\
  \bottomrule
\end{tabular}
\end{table}

\begin{figure}[tbp]
  \centering
  \includegraphics{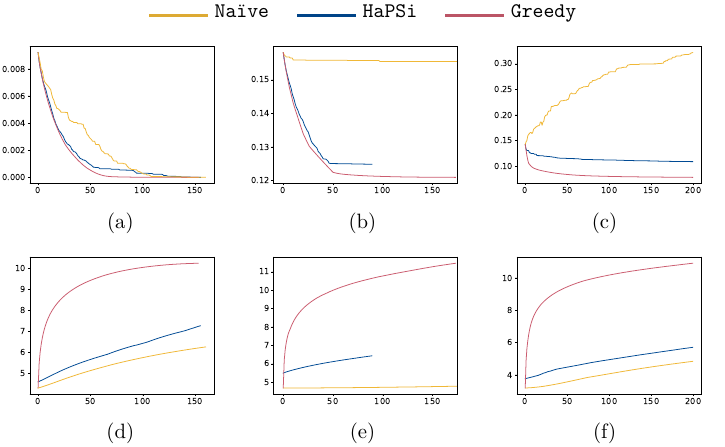}
    \caption{\textbf{Top row:} Relative reconstruction error on the $y$-axis and number of tiles on the $x$-axis. (a) Tiles are redescriptions from the \wcmam data. (b) Tiles are frequent itemsets from the \dialect data. (c) Tiles are rank-1 Boolean matrices for BMF from the \dialect data. \textbf{Bottom row:} Time (in seconds) on $y$-axis and number of tiles on $x$ axis. (d)--(f) as (a)--(c).}
    \label{fig:real}
  \end{figure}

\paragraph{Redescription mining.}
We used redescriptions mined using the fast redescription mining algorithm Fier~\cite{karjalainen2024fast}. These results contained \num{161} redescriptions, some of them very similar to each other. The results are shown in Fig.~\ref{fig:real}(a) and (d).
The results show that \ownalgo is able to find very fast a good set of approximately 50 redescriptions. It should be noted that this data can be fully covered using all \num{161} tiles, so all methods converge towards the end. Notice also that \greedy is very slow.

\paragraph{Tiling databases.}
We mine frequent itemsets from the \dialect dataset with minimum frequency of \qty{15}{\percent}. This gave us \num{28535} frequent itemsets. We used all of these as tiles, and searched for a set of maximum \num{1000} tiles. The results are in Fig.~\ref{fig:real}(b) and (e). We see that \ownalgo is again very good up to about \num{50} itemsets, after which it has converged. It is again significantly faster than \greedy and significantly better than \naive.

\paragraph{Boolean matrix factorization.}
We created the rank-1 Boolean matrices using the association matrix technique used by the Asso algorithm and the restarted random walks technique~\cite{miettinen2020}. These rank-1 matrices can cover $0$s in the data, unlike in the previous cases. The results using the \dialect data are shown in Fig.~\ref{fig:real}(c) and (f). The results are similar to the other real-world use cases, except that in this case the \naive algorithm increases its error consistently. But again \ownalgo is competitive with \greedy, especially when we consider the significant advance it has on running time. 

We also ran experiments with tiles found only by the association matrix technique using the two large corpus matrices. This created approximately \num{5000} tiles for \newsgroups and \num{12000} for \abstracts. \ownalgo took \qty{850}{\second} for \abstracts, while \greedy took \qty{101}{\hour}. For \newsgroups \ownalgo took \qty{665}{\second} and \greedy \qty{74}{\hour}. This shows that \ownalgo can find results efficiently with input sizes where \greedy is unable to produce the results in a reasonable amount of time.

\section{Related Work}
\label{sec:related-work}

Tiling databases was first studied in~\cite{geerts2004a}, where also the use of the greedy algorithm was proposed. The problem was later independently studied in~\cite{xiang2010}. The tiling was later extended to hierarchical representations (where tiles contain other tiles)~\cite{tatti2012}, data streams~\cite{lam2014}, and ranked data~\cite{levan2014}, among others.

Boolean matrix factorization (BMF) can be seen as a variant of tiling. It has been originally studied in combinatorics (see~\cite{monson1995} and references therein), but it has also seen significant research interest in data mining (see~\cite{miettinen2020} and references therein). An exact Boolean matrix factorization is the same as exact tiling, and~\cite{belohlavek2010a} propose an algorithm for exact BMF based on greedy Set Cover algorithm.

The aforementioned methods use reconstruction error as the primary means of deciding which tiles or rank-1 matrices to use. Minimum Description Length (MDL) principle is another popular method for selecting the pattern set (see, e.g.~\cite{vreeken2011,mampaey2011,miettinen2014b}). In MDL, the idea is to select those tiles that compress the data the best.

Another method for selecting the set of patterns was proposed by De Bie~\cite{debie2011}. His proposal is based on modelling the data with a maximum-entropy distribution, and selecting the tiles that are the most surprising (i.e.\ least likely) under this distribution. After each selected tile, the distribution is updated by constraining it with the selected tile such that the new distribution has the maximum entropy over all those distributions where the found tiles appear. Later, \cite{kalofolias2018} extended that approach to selecting a set of redescriptions.

Redescription mining~\cite{galbrun2018a} was proposed in~\cite{ramakrishnan2004} and has seen various algorithms and applications proposed since (see~\cite{galbrun2018a}). The first redescription set mining approach was proposed in~\cite{galbrun2018b} and later \cite{kalofolias2018} proposed a version based on subjective interestingness.

The idea of using bottom-$k$ hashing for size estimation was first proposed in~\cite{bar2002counting}, while~\cite{amossen14better} extended it to the size of Boolean matrix product and explained the efficient algorithm we use as the basis of \ownalgo.

\section{Conclusions}
\label{sec:conclusions}

Bottom-$k$ hashing can provide significant speedups for selecting a good set of patterns with minimal effects on the quality of patterns selected. We showed how our algorithm can be applied to tiling, Boolean matrix factorization, and redescription set mining, but we believe that there are other pattern set mining problems where our approach can be useful, as well.

The hashing techniques presented in this paper are limited to
reconstruction error. Algorithms like \ownalgo
are not directly good for selecting itemsets based on measures like MDL,
as they pay no attention to the ``cost'' of the selected
patterns. Indeed, we can see in Fig.~\ref{fig:real} that \ownalgo
sometimes selects patterns that improve the error very little if at
all; such patterns are very unsuitable for optimizing MDL, say. In
principle, the description length could be taken into account when
selecting the patterns, but it is not clear if the bottom-$k$ hashing
would ever find patterns that are good in MDL-sense. Developing these
algorithms is an interesting topic for a future work.

Being able to select the pattern set faster can also facilitate novel
algorithms that build sets of pattern sets. Such algorithms can
provide a hierarchical view to the data, supporting exploration with
few high-level pattern-sets-as-patterns, each of which can be ``zoomed
in'' to see the actual patterns. 

Our method does not allow  novel unethical uses of data mining
techniques although the vast speed improvements do allow existing
potentially unethical uses to be scaled to much larger data sets.

\bibliographystyle{splncs04}

\end{document}